\documentstyle[12pt]{article}

\begin{document}
\thispagestyle{empty}
\begin{center}
{\Large\bf
   Geometrical action for $w_\infty$ algebra as a reduced symplectic
                        Chern-Simons theory}
\end{center}
\bigskip
\bigskip
\bigskip
\bigskip
\begin{center}
{\large\bf
               R.P. Manvelyan and R.L. Mkrtchyan }
	       {\footnote{E-mail: mrl@dircom.erphy.armenia.su}}\\
\bigskip
{\it
	       Yerevan Physics Institute \\ Yerevan,Alikhanyan Br.st.2,375036
	       Armenia}\\
\end{center}
\bigskip
\bigskip
\bigskip
\bigskip
\bigskip
\begin{abstract}
 The geometric action on a certain
orbit of the group of the area-preserving  diffeomorphisms
is considered, and it  is shown,  that  it  coincides  with  a
special reduction of   the three-dimensional Chern-Simons
theory, under which group and space coordinates are identified.
\end{abstract}
\setcounter{page}0
\newpage

\section{Introduction}
     The method of coadjoint orbits proves to be  useful  in  many
problems in physics and mathematics [1]. The area of  applications
of that method is enlarged in the last years  by  considering  the
coadjoint orbits of infinite-dimensional groups and algebras, such
as Virasoro, Kac-Moody, diffeomorphisms, etc [2].  In  the  present
work  we  would  like  to  consider  the  previously   constructed
geometric action on the coadjoint  orbits  of  the  group  of  the
area-preserving diffeomorphisms of 2d surfaces [3], for a  special
orbit,   and   will   find   an   interesting   connection    with
three-dimensional Chern-Simons theory [4].

\section{Geometrical and Chern-Simons action}

     The general form of a Hamiltonian system on a coadjoint orbit
of the  (non  centrally-extended)  group  of  the  area-preserving
diffeomorphisms   of   some   two-dimensional   surface   is   the
following [3]:
$$     S_{geom}(F)=\int d^2xdt[ B_i (F^j)dF^i /dt + \lambda (\{F^1
,F^2 \}-1) - H(F)] \eqno (1)$$
 where $B_i (F(x,t))$ is the
one-form, which characterizes the orbit, the $F^i$ are the group
elements, parametryzing  orbit, and  $\{,\}$ is the Poisson
bracket, inverse to closed two-form, which in the simplest case
of torus or plane is $dx^1 dx^2$ .  The $F^i$ have to satisfy
the $\{F^1 ,F^2\}=1$ restriction of the area-preservation,
which is taken into account in the second term  in  (1) through
the Lagrange multiplier $\lambda(x,t)$. The first term  in  (1) is
the Kirillov-Kostant symplectic one-form [2], the last is some
Hamiltonian.  We would like to set $H=0$,choose the special
orbit, corresponding to $$ B_i(F) = \epsilon_{ij}F^j $$ and make the
following field redefinition: $$F^i = x^i + \epsilon^{ij} a_j,$$
$$\lambda = a_0$$ Then S receives a form (after rescaling of
 $a = a_\mu dx^{\mu}$ , $\mu=0,1,2$ )
$$  S(a) = k\int Tr [ada + (1/3)a\{a,a\}] \eqno (2)$$
 from which it is evident, that $S$ possess the gauge invariance
 with respect to
the following gauge transformation with  parameter $\varepsilon (x)$:
 $$\delta a_\mu = \partial_{\mu} \varepsilon + \{a_{\mu} ,\varepsilon\}\eqno
(3)$$ That follows from the fact, that the theory (2) has the form, very
similar to the standard Chern-Simons [4]: $$ S = k\int AdA +
(1/3)A[A,A]$$ theory, with usual Lie  algebra commutator of
potentials $A$ replaced by the Poisson bracket $\{,\},$ which also
satisfies  the Jacoby identity and actually may be considered as
a commutator  in a Lie algebra of a functions on a surface, or
as a  commutator  in the Lie algebra of  the area-preserving
diffeomorphisms  of a surface.  The only,  but essential,
difference  with usual Chern-Simons theory is in the following.
In the standard situation the Lie algebra indexes of a
vector-potential $A^a_\mu$ , i.e.  index  a, are completely different
from its space-time "indexes",  i.e.  its $x$ dependence. In (2)
these two sets of indexes  are  identified:  the Poisson bracket
contains the space derivatives $\partial / \partial x^i$ . We shall show now,
that the action (2) may be obtained as a some reduction of  the symplectic
Chern-Simons theory, and this reduction  really means an identification of the
space and group coordinates.  Let's consider the so-called symplectic
Chern-Simons  theory, i.e. the CS theory with gauge  group $w_\infty$  of  the
     area-preserving transformations of some $2d$ surface with coordinates
$y^i$ ,$ i=1,2.$ We shall take that  $2d$  surface  to be  a  torus,  which
simplifies formulae,  although  the general  case  may  be
considered.  The representation of the gauge algebra, in which
the gauge potentials $A$  receive their values, is the following:
$A $ is the function  not only of space-time coordinates $x$ , but
also  of  torus  coordinate $y^i$ , $A =A_{\mu}(x,y)dx^{\mu}$. The commutator
is the Poisson bracket:  $$ [A ,A ] = \{A ,A\} =
\epsilon^{ij}(\partial/\partial y^i)A(\partial/\partial y^j)A \eqno(4) $$ and
 the trace is the integral over $y$ coordinates, so the  action and gauge
transformations of  the  symplectic  CS theory  may  be written as $$ S_{CS} =
\alpha\int (d^2y)(AdA + (1/3)A\{A,A\})\eqno(5)$$ $$ \delta A = d\varepsilon +
\{A ,\varepsilon\}\eqno(6)$$ which  is  similar  to   (2),   but   different,
due to   the, first,integration over torus and second,  the  fact that
Poisson bracket now have to be taken with respect  to  the $y$ - coordinates.
Now we would like to adopt a following limiting procedure,  which gives a
certain reduction of $S$  . We make an anzatz $$A_{\mu} (x,y) = a_{\mu} (x^{0}
,(x^{i} +y^{i} )/2) exp((x^{i} -y^{i})^{2} /2t)\eqno(7)$$ $$\alpha = k/t$$ and
take the limit $t\Rightarrow{0}$.  Substituting (7) into  (5), we have, taking
 a limit $t\Rightarrow{0}$, using $$(1/t^{2} )exp((x^{i} -y^{i})^{2}/2t)
\Rightarrow\delta^{(2)} (x-y), $$ and rescaling $\alpha$ and $k$:  $$S_{CS}
\Rightarrow{S}$$

\section{Observables}

     The standard set of observables for CS theories consist  from
an arbitrary Wilson loops [4]. The definition of the  Wilson  loop
for our theory cannot be done straightforwardly, since  the  group
and  space  coordinates  are  mixed  (identified)   and   standard
definitions fail. The way  out  is  in  the  following  artificial
disentangling of the space and group coordinates. The gauge  group
of our theory is the group of the area-preserving transformations.
Usually the infinitesimal group element, corresponding to  element
$dx^{\mu}$ of the curve $C$, is given by the  expression $1+dx^{\mu} A_{\mu}$.
In  ourcase  this  element  have  to  be  infinitesimal   area-preserving
diffeomorphism. This follows from the rule (3), (6) for the  gauge
transformations. For the symplectic $CS$ theory (5)  the  analog  of
$1+dx^{\mu} A_{\mu}$  is, clearly, the infinitesimal diffeomorphism of torus
  $$  y^i \Rightarrow {y^i  + \epsilon^{ij} dx^{\mu}( \partial/ \partial
y^{j})A_{\mu} (x,y)} \eqno(8) $$ and to each oriented curve $C$ with given
initial and final points $x_1$  and $x_2$   corresponds an element  of   the
group of the area-preserving diffeomorphism  of the torus,  given  by   the
composition of the infinitesimal diffeomorphism (8).  As  usual, the
corresponding group element, and particularly (8), transforms homogeneously
under gauge transformation, which is  given  by  the local ($x$-dependent)
area-preserving  $F_{x} (y)$:
$$ y^i +\epsilon^{ij} dx^{\mu} (\partial/\partial y^{j})A_{\mu} (x,y)
\Rightarrow $$
  $$(F_{x}^{i})^{-1} (F_{x+dx}^{i}(y)
+\epsilon^{ij} dx^{\mu}(\partial /\partial y^{j}) A_{\mu}(x,F_{x+dx}^{i}(y)))
\quad\eqno (9)$$
 It is easy  to check,  that   for the infinitesimal gauge
transformation $ F_{x}^{i}(y) = y^{i} +\epsilon^{ij}(\partial/\partial y^{j})
      \varepsilon(x,y)$ Eq.(9) is really equivalent to (5).  Evidently, such a
     formulae are absent for the reduced $CS$ action (2), so, as was mentioned
     above, we apply the following trick.  Let's consider the finite gauge
     transformation (3)  for (2), given by the functions $F^{i}(x)$ with
     area-preserving condition $ det\frac{\partial{F}}{\partial{x}} =1 $, and
  introduce the following functions, which formally are the special gauge
     transformations for the symplectic $CS$ theory (5):
$$   F_{x}^{i}(y) =  F^{i}(x+y)-x^{i} \eqno (10) $$
Here (and below) $x+y$ means $(x^{0},x^{i} + y^{i})$.  It is easy to check the
important property, that composition of two
gauge transformations  $F^{1}$ and $F^{2}$ is equivalent to  the composition
of $ F_{x}^1$  and $F_{x}^2 $:
$$  F_{x}^{1} \circ F_{x}^{2} (y) =F_{x}^{1}(F_{x}^{2} (y))$$
$$ = F_{x}^{1} (F^{2} (x+y) - x) = F^{1} (F^{2} (x+y) - x + x) - x = (F^{1}
\circ F^{2} )_{x}
(y) \eqno (11) $$
Similarly, we built from $a_{\mu} (x)$ the $y$-dependent
 $A_{\mu} (x,y)$:  $$    A_{\mu} (x,y) = a_{\mu} (x+y) \eqno (12) $$ and the
gauge transformation of $A_{\mu} (x,y)$ derived from definition (12) and gauge
transformation of $a_{\mu} (y)$, appears to be the same in form as in $CS$
(5), with gauge parameter $ \varepsilon_{x}(y) = \varepsilon (x+y)$, which
is an infinitesimal form of finite gauge transformation $$  F_{x}^{i}
(y)=F^{i} (x+y)-x^{i} \approx (x^i + y^i + \epsilon^{ij} (\partial/\partial
y^{j} )\varepsilon (x+y) - x^i $$
$$ = y^i + \epsilon^{ij} (\partial/\partial y^{j} )\varepsilon(x+y) \eqno(13)
$$ Then we construct an $y$ -diffeomorphism
corresponding to element $dx^{\mu}$ of the curve $C$ by the same expression
(8), and the main statement is that the transformation (9) with definition
 (10) again  is equivalent to the gauge transformation of $ a_{\mu} (x)$,
given by the (3).  So we can proceed further in a standard way, taking the
group elements for a curves $C$ with coinciding initial and final points, and
taking the character of that element.  Important is the problem of defining
the  symmetries  of the action (2). One of the symmetries is the gauge
symmetry  (3).  The other one is the invariance with respect  to the arbitrary
reparametrization of time $x^{0} \Rightarrow f(x^{0})$, under which $a_{\mu}$
behaves  as a vector. Also, action (2) is invariant w.r.t.  the time-dependent
area-preserving reparametrizations of $x^{1}$ and $x^{2}$ , under which $a$
transforms as a one-form.  The investigation of (2) in analogy with usual $CS$
theory remains an open problem to be solved.
 \section{Acknowledgments}
      Authors would like  to  thank  M.Vasiliev   and  R.Flume  for
discussions.
     This work was supported in part by the grant 211-5291 YPI  of
the  German  Bundesministerium  fur  Forschnung  und  Technologie,
Federal Republic of Germany.

 \end{document}